\documentclass[12pt]{iopart}
\usepackage{iopams}
\usepackage[dvips]{graphicx}
\begin{document}

\title[Magnetization plateaux of Cs$_2$CuBr$_4$]
{Magnetization plateaux of S = 1/2 two-dimensional frustrated antiferromagnet Cs$_2$CuBr$_4$}

\author{
	T. Ono\dag\footnote[3]{E-mail: o-toshio@lee.phys.titech.ac.jp}, H. Tanaka\dag, O. Kolomiyets$^1$, H. Mitamura$^1$, T. Goto$^1$, K. Nakajima$^2$, A. Oosawa$^3$, Y. Koike$^3$, K. Kakurai$^3$, J. Klenke$^4$, P. Smeibidle$^4$, M. Mei{\ss}ner$^4$
	}

\address{
\dag\ Dept. of Phys., Tokyo Institute of Technology, 152-8551 Tokyo, Japan
}

\address{
$^1$\ Institute for Solid State Physics, The University of Tokyo, Chiba 277-8581, Japan
}

\address{
$^2$\ Neutron Scattering Laboratory, Institute for Solid State Physics, University of Tokyo, Ibaraki 319-1106, Japan
}

\address{
$^3$\ Advanced Scientific Research Center, Japan Atomic Research Institute, Ibaraki 319-1195, Japan
}

\address{
$^4$\ Hahn-Meitner-Institut, Glienicker Stra{\ss}e 100, D-14109 Berlin, Germany
}

\begin{abstract}
The field induced magnetic phase transitions of Cs$_2$CuBr$_4$ were investigated by means of magnetization process and neutron scattering experiments. This system undergoes magnetic phase transition at Ne\'{e}l temperature $T_\mathrm{N}=1.4$\,K at zero field, and exhibits the magnetization plateau at approximately one third of the saturation magnetization for the field directions $H\parallel b$ and $H\parallel c$. In the present study, additional symptom of the two-third magnetization plateau was found in the field derivative of the magnetization process. The magnetic structure was found to be incommensurate with the ordering vector $\boldsymbol{Q}=(0, 0.575, 0)$ at zero field. With increasing magnetic field parallel to the $c$-axis, the ordering vector increases continuously and is locked at $\boldsymbol{Q}=(0, 0.662, 0)$ in the plateau field range $13.1\,\mathrm{T} < H < 14.4\,\mathrm{T}$. This indicates that the collinear \textit{up-up-down} spin structure is stabilized by quantum fluctuation at the magnetization plateau.
\end{abstract}



\maketitle

\section{Introduction}

In the last two decades, triangular antiferromagnetic (TAF) system was studied extensively using hexagonal ABX$_3$-type antiferromagnets and many types of the phase transitions due to the frustration were found\,\cite{Collins}. Within the classical model, the spin structure of the two dimensional Heisenberg TAF in a magnetic field cannot be determined uniquely, because the number of the parameters those are given by the equilibrium conditions are insufficient for the determination of the spin configuration, and thus there remains ``non-trivial continuous degeneracy'' at the ground state\,\cite{Chubukov}.\par
\begin{figure}[tbp]
	\begin{minipage}{0.45\linewidth}
		\begin{center}
			\includegraphics[width=\linewidth,clip]{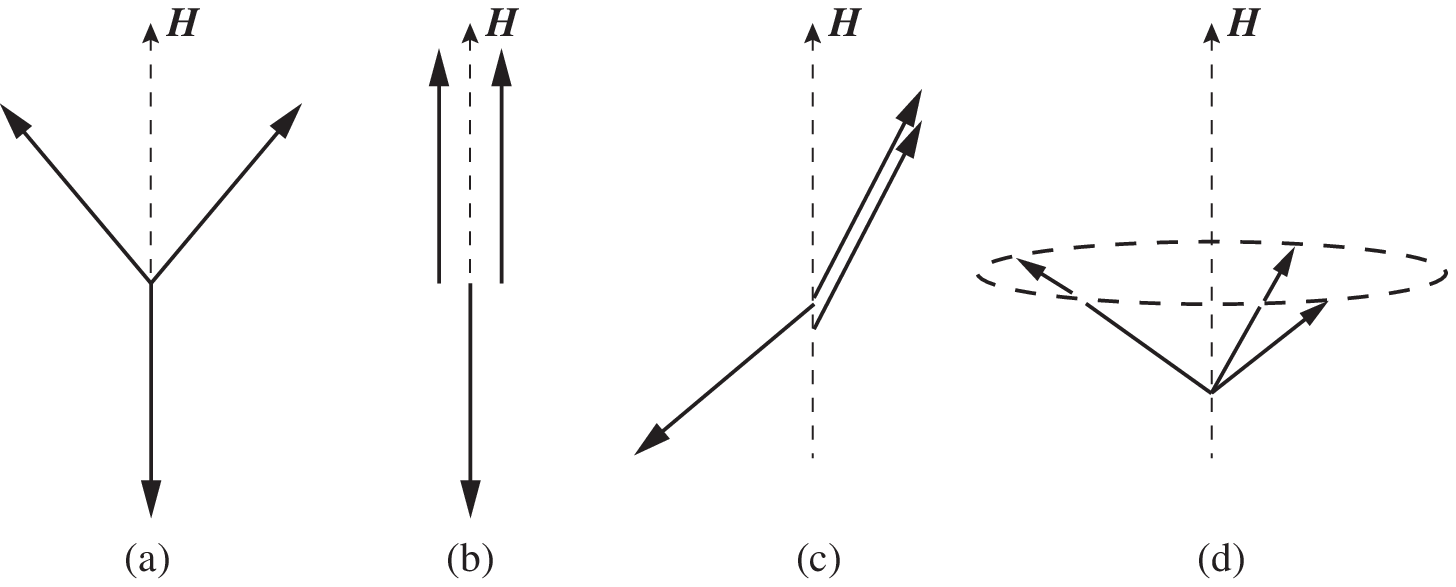}
		\end{center}
		\caption{Spin structure of 2D TAF in the magnetic field. (a) Low-field coplanar structure, (b) \textit{up-up-down} structure, (c) high-field coplanar structure and (d) umbrella structure.}
		\label{structures}
	\end{minipage}
	\hspace{5mm}
	\begin{minipage}{0.45\linewidth}
		\begin{center}
			\includegraphics[width=0.7\linewidth,clip]{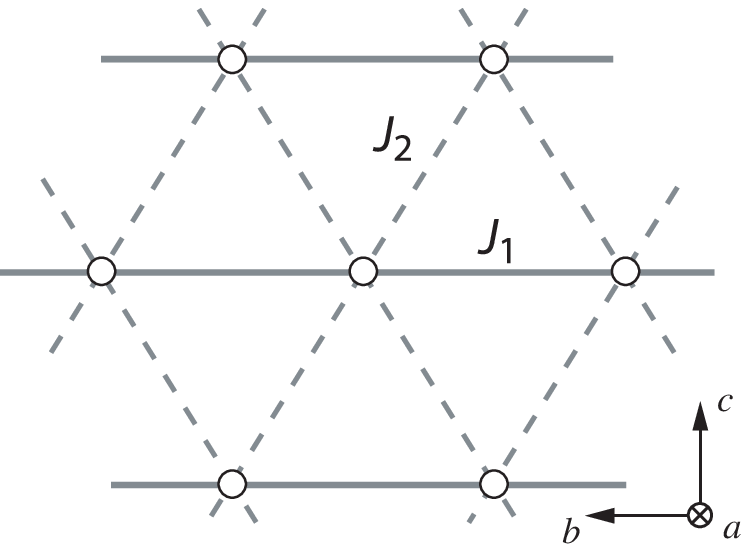}
		\end{center}
		\caption{Antiferromagnetic interactions $J_1$ and $J_2$ within the \textit{bc}-plane. The open circles denote Cu$^{2+}$-ions.}
		\label{bc_lattice}
	\end{minipage}
\end{figure}

Chubukov and Golosov\,\cite{Chubukov} suggested that quantum fluctuation can lift the continuous degeneracy at the ground state. They argued that the Heisenberg TAF undergoes a successive phase transition as shown in Fig.\,\ref{structures} (a), (b) and (c) in that order with increasing magnetic field. 
The intermediate collinear spin structure is stabilized in a finite field region and the magnetization plateau arises at one-third of the saturation magnetization.\par
The magnetic phase transition of TAF induced by the quantum fluctuation was firstly found in CsCuCl$_3$\,\cite{Nojiri,Nikuni}. Due to the weak easy-plane anisotropy, the umbrella spin structure as shown in Fig.\,\ref{structures} (d) is realized in the low-field region for $H\parallel c$-axis, and the spin structure such as (a) and (b) are hidden by the magnetic anisotropy. Consequently that CsCuCl$_3$ undergoes the field induced phase transition from the structure (d) to (c), which is accompanied by the small magnetization ``jump''\,\cite{Nojiri}.\par
In Cs$_2$CuBr$_4$, the arrangement of $S=1/2$ Cu$^{2+}$-ion can be regarded as the distorted TAF in the \textit{bc}-plane as shown in Fig.\,\ref{bc_lattice}\,\cite{Tanaka,Ono}. The magnetic properties of the isostructural Cs$_2$CuCl$_4$ have been investigated extensively by Coldea \etal\,\cite{Coldea1,Coldea2,Coldea3,Coldea4}, and Cs$_2$CuCl$_4$ was characterized as the 2-dimensional (2D) TAF. Therefore, it is expected that Cs$_2$CuBr$_4$ is also described by an $S=1/2$ 2D TAF.\par
In our previous measurements for Cs$_2$CuBr$_4$\,\cite{Tanaka,Ono}, the magnetic ordering was found at $T_\mathrm{N}=1.4$\,K, which is more than twice as large as $T_\mathrm{N}=0.62$\,K in Cs$_2$CuCl$_4$\,\cite{Coldea1}. The notable feature of the present system is the magnetization plateau at one-third of the saturation magnetization $M_\mathrm{S}$ observed for the magnetic field $H\parallel b$ and the \textit{c}-axes in the ordered phase as shown in below. Since the magnetization plateau is observed for two different field directions, its origin is attributed not to the uniaxial magnetic anisotropy but to the quantum effect. On the other hand, no plateau is observed for Cs$_2$CuCl$_4$. To the best of our knowledge, the magnetization plateau observed in Cs$_2$CuBr$_4$ is the first experimental example of the one-third plateau stabilized by quantum fluctuation in TAF. In the present study, we have remeasured the magnetization process in order to improve the signal-to-noise ratio and performed the elastic neutron scattering experiments in order to determine the magnetic structure of Cs$_2$CuBr$_4$ in the field parallel to the \textit{c}-axis up to the plateau region.

\section{Experiments}
Single crystals of Cs$_2$CuBr$_4$ were grown by the slow evaporation method from the aqueous solution of CsBr and CuBr$_2$ in the mole ratio 2 : 1.
The high-field magnetization measurement was performed using an induction method with a multilayer pulse magnet at the Ultra-High Magnetic Field Laboratory, Institute for Solid State Physics, The University of Tokyo. Magnetization data were collected at $T=400$\,mK in magnetic fields up to 35 T.\par
The elastic neutron scattering experiments at zero field were performed at HER and LTAS triple axis spectrometer installed at JAERI, Tokai. A single crystal with $\sim$3\,cm$^3$  was used for this measurement. \par
The field dependence of the elastic neutron scattering experiments were performed at E1  triple axis spectrometer installed at experimental reactor of Hahn-Meitner-Institute, Berlin. The sample used was approximately 1\,g in mass. Sample was mounted on the  sample stage of the dilution stick DS-X and the dilution stick was loaded into VM-1 14.5\,T vertical superconducting magnet. Measurements were performed for the $(a, b)$ horizontal scattering plane. \par

\section{Results and Discussion}
\subsection{Magnetization process}
\begin{figure}[tbp]
	\begin{center}
		\includegraphics[width=0.72\linewidth,clip]{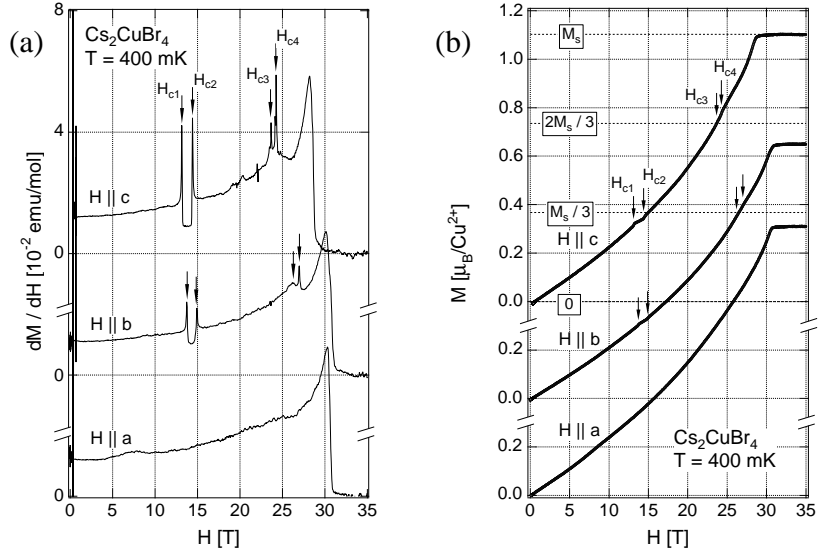}
	\end{center}
	\caption{(a) Field derivative of the magnetization and (b) the magnetization as a function of the external field applied for $H\parallel a$-, $b$- and the $c$-axis. Measurements have been done at $T=400$\,mK.}
	\label{magnetization}
\end{figure}

	Figure\,\ref{magnetization} shows the field dependences of $\mathrm{d}M/\mathrm{d}H$ and the magnetization of Cs$_2$CuBr$_4$ for the field direction $H$ parallel to the $a$-, $b$- and $c$-axis measured at $T=400$\,mK. 
The data were taken in sweeping down magnetic field. The magnetization saturates at $H_\mathrm{S}\approx 30$\,T. The difference between the absolute values of the saturation fields and the saturation magnetizations for the three different field directions should be due to the anisotropy of \textit{g}-factor. For $H\parallel a$, $\mathrm{d}M/\mathrm{d}H$ does not have any anomaly up to the saturation field. On the other hand, for $H\parallel b$ and $H\parallel c$, $\mathrm{d}M/\mathrm{d}H$ show the several sharp peaks indicated by the arrows $H_\mathrm{c1}\sim H_\mathrm{c4}$ in figure\,\ref{magnetization} (a). The anomalies $H_\mathrm{c1}$ and $H_\mathrm{c2}$ correspond to $1/3$ magnetization plateau. The level of the $1/3$ magnetization plateau is slightly lower than one third of the saturation magnetization $M_\mathrm{S}/3$.  A couple of peaks labeled as $H_\mathrm{c3}$ and $H_\mathrm{c4}$ which are seen for $H\parallel b$ and $H\parallel c$ also has the feature of the magnetization plateau. At  $H_\mathrm{c3}$ and $H_\mathrm{c4}$, the magnetization is just above $2M_\mathrm{S}/3$. For the magnetic system that does not have uniaxial anisotropy, the appearance of the magnetization plateau can be attributed to the existence of a quantum excitation energy gap. Since the distance of the field between these two peaks $H_\mathrm{c3}$ and  $H_\mathrm{c4}$ is approximately 0.6\,T, the energy gap around 2/3 plateau $\Delta_{\frac{2}{3}}$ is roughly estimated as $\Delta_{\frac{2}{3}}/k_\mathrm{B}\sim 0.4$\,K which is comparable with the measuring temperature. Therefore, the finite temperature effect cannot be neglected for 2/3-magnetization plateau. In order to confirm the existence of the 2/3-magnetization plateau, the measurements with the lower temperature is in progress.

\subsection{Neutron scattering}
\begin{figure}[tbp]
	\begin{center}
		\includegraphics[width=0.8\linewidth,clip]{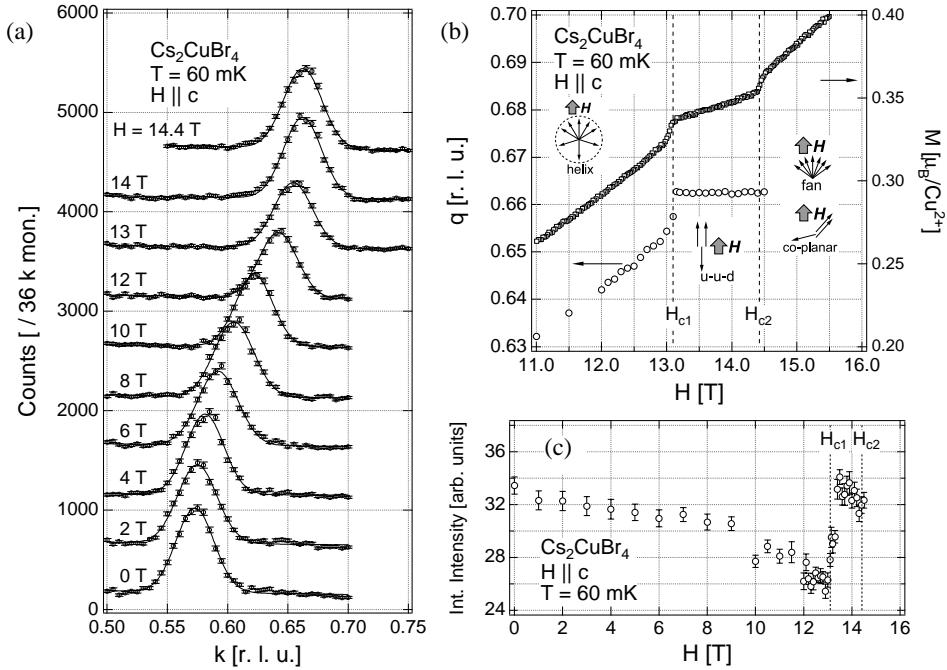}
	\end{center}
	\caption{(a) Magnetic Bragg reflections scanned along $\boldsymbol{Q}=(0, k, 0)$ at various external fields. (b) The field dependence of the ordering vector $\boldsymbol{Q}=(0, q, 0)$ (left abscissa) and the magnetization process (right abscissa) around the plateau region. The magnetization curve was measured at $T\sim 400$\, mK. (c) The field variation of the integrated intensities of the primary magnetic Bragg peak.}
	\label{profiles}
\end{figure}

Figure\,\ref{profiles} (a) shows the typical scan profile along the \textit{b}*-direction around at $T=60$\,mK. 
Magnetic Bragg reflection appears at $\boldsymbol{Q}_0=(0, 0.575, 0)$. This result indicates that the magnetic structure is incommensurate with the lattice along the \textit{b}-axis. The origin of the incommensuration is attributed to the competition between $J_1$ and $J_2$ (see Fig.\,\ref{bc_lattice}). The ordering vector for Cs$_2$CuCl$_4$ is $\boldsymbol{Q}_0=(0, 0.528, 0)$\,\cite{Coldea1}. In the classical theory, the ordering vector $\boldsymbol{Q}_0=(0, q, 0)$ is given by $\cos(\pi q)=-J_2/(2J_1)$. Using this equation, we obtain $J_2/J_1=0.467$ for Cs$_2$CuBr$_4$ and $J_2/J_1=0.175$ for Cs$_2$CuCl$_4$. This result implies that Cs$_2$CuBr$_4$ is more frustrated than Cs$_2$CuCl$_4$. Of course, since quantum effects cannot be neglected in the present system, we should determine the exchange interactions from the magnetic excitation.\par
With increasing magnetic field parallel to the \textit{c}-axis, the value of the ordering vector $q$ increases continuously as shown in Fig.\,\ref{profiles} (b). On the other hand, the integrated intensity decreases gradually up to the transition field $H_{\mathrm{c1}}$ and shows the jump just after the transition field $H_{\mathrm{c1}}$ as shown in Fig.\,\ref{profiles} (c). The neutron scattering cross section reflects the spin component perpendicular to the scattering vector. Since the magnetization process for $H\parallel b$ and $H\parallel c$ are almost the same, it can be deduced that the magnetic moments almost lie in the $bc$-plane as observed in Cs$_2$CuCl$_4$\,\cite{Coldea1}. Thus, the integrated intensity is expected to be increased with increasing external field, because of the increase of the spin component parallel to the \textit{c}-axis. The reason for the decrease of the intensity for $H<H_\mathrm{c1}$ is unclear.\par
Figure\,\ref{profiles} (b) shows the external field dependence of the ordering vector $\boldsymbol{Q}_0=(0, q, 0)$ measured for $H\parallel c$-axis. The value of the ordering vector \textit{q} shows a steep increase just below the lower edge field $H_{\mathrm{c1}}=13.15$\,T, and is locked at $q=0.662$ above $H>H_\mathrm{c1}$. In Fig.\,\ref{profiles} (b), the magnetization curve measured at $T\sim 400$\,mK (right abscissa) is also plotted for comparison. The values of $H_\mathrm{c1}$ determined from the magnetization and the present neutron scattering measurements show the excellent agreement. This fact implies obviously that the three-sublattice \textit{up-up-down} spin structure parallel to the external field is realized at the plateau. Measurements have been done up to $H=14.5$ T, however, any evidence of the magnetic phase transition at $H=H_\mathrm{c2}$ could not be observed. The transition field $H_\mathrm{c2}$ seems to be just above the field range of the present measurement. \par
For the classical helical spin system, a transition from a helical spin structure to a fan structure (see the illustration in Fig.\,\ref{profiles} (b)) can occur when an external field is applied in the easy plane\,\cite{Nagamiya}. The magnetization process of RbCuCl$_3$ which is described as ferromagnetically stacked distorted TAF shows the \textit{helix-fan} transition for the field parallel to the triangular plane\,\cite{Maruyama,Jacobs}. For RbCuCl$_3$, the quantum fluctuation should be suppressed by the three dimensional interaction. There is possibility that the fan spin structure is also realized in Cs$_2$CuBr$_4$ for the field region around the saturation field. The fan structure is characterized by the two ordering components parallel and perpendicular to the field. Thus, the Bragg reflections should split from $q=0.662$, if the fan structure was realized in the high field region $H>H_\mathrm{c2}$. If the high-field coplanar structure that has an advantage of quantum fluctuation (shown in Fig.\,\ref{structures} (b) ) was realized in the field region $H>H_\mathrm{c2}$, the ordering vector does not move from $q=0.662$. In order to confirm the spin structure in the field range $H>H_\mathrm{c2}$, the measurements in more high fields are necessary.\par

\section{Conclusions}
We have presented the results of the magnetization process and the elastic neutron scattering experiments on Cs$_2$CuBr$_4$. It was found that the magnetization curves have the plateaux at approximately 1/3 and 2/3 of the saturation magnetization. At zero field, the spin structure in the ordered phase is incommensurate with the ordering vector $\boldsymbol{Q}_0=(0, q, 0)$ with $q=0.575$. With increasing field parallel to the \textit{c}-axis, the value of $q$ is increased, and then locked at approximately $q=2/3$ above the lower edge field $H_\mathrm{c1}=13.15$\,T of the magnetization plateau. Thus it can be concluded that the collinear (\textit{up-up-down}) spin structure is realized in the 1/3 plateau field region. This collinear spin structure is stabilized by the quantum fluctuation. The spin structures in the higher field region are unclear at present.

\end{document}